# From Data to Alloys: Predicting and Screening High-Entropy Alloys for High Hardness Using Machine Learning

Rahul Bouri[1], Manikantan R. Nair[1], Tribeni Roy[1]

[1]Department of Mechanical Engineering, Birla Institute of Technology and Science Pilani, Rajasthan, 333031, India

**Abstract**

The growing need for structural materials with strength, mechanical stability, and durability in extreme environments is driving the development of high-entropy alloys. These are materials with near-equiatomic mixing of five or more principal elements, and such compositional complexity often leads to improvements in mechanical properties and high thermal stability, etc. Thus, high-entropy alloys have found their applications in domains like aerospace, biomedical, energy storage, catalysis, electronics, etc. However, the vast compositional design and experimental exploration of high-entropy alloys are both time-consuming and expensive and require a large number of resources. Machine learning techniques have thus become essential for accelerating high-entropy alloys discovery using data-driven predictions of promising alloy combinations and their properties. Hence, this work employs a machine learning framework that predicts high-entropy alloy hardness from elemental descriptors such as atomic radius, valence electron count, bond strength, etc. Machine learning regression models, like LightGBM, Gradient Boosting Regressor, and Transformer encoder, were trained on experimental data. Additionally, a language model was also fine-tuned to predict hardness from elemental descriptor strings. The results indicate that LightGBM has better accuracy in predicting the hardness of high-entropy alloys compared to other models used in this study. Further, a combinatorial technique was used to generate over 9 million virtual high-entropy alloy candidates, and the trained machine learning models were used to predict their hardness. From screening these combinatorially generated high-entropy alloys, it was found that compositions, such as $Al_{68.4}Ni_{13.5}Zn_{13.4}Re_{4.7}$, $Al_{69.0}Fe_{10.9}Mo_{8.4}Zn_{11.8}$, and $Al_{69.6}Co_{12.9}Mo_{7.1}Zn_{10.4}$, have a high Vickers' hardness of 600 (Hv). This study shows how machine learning-driven high-throughput screening and language modelling approaches can accelerate the development of next-generation high-entropy alloys.

1. **Introduction**

High-entropy alloys (HEAs) have shown their effectiveness in demanding applications where conventional alloys often fail due to thermal or mechanical degradation, such as aerospace engine components that require high creep resistance at elevated temperatures, nuclear energy systems that demand durability under radiation, and automotive systems that benefit from high strength-to-weight ratios. HEAs are mostly composed of five or more principal elements mixed in near-equiatomic proportions, which leads to high configurational entropy [1]. This entropy stabilizes simple solid-solution phases and suppresses the formation of complex intermetallic compounds [2]. As a result, these materials exhibit superior mechanical properties, like high strength, corrosion and wear resistance, fracture toughness, and thermal stability. Such features make HEAs suitable for extreme operating environments across aerospace, automotive, defense, and energy sectors [3]. The growing industrial adoption of HEAs highlights the need for efficient discovery strategies capable of navigating the vast compositional design space. However, identifying novel compositions with tailor-made properties is challenging due to the high compositional complexity, elemental interactions, and variations created by processing techniques.

Traditional experimental approaches for HEA design are often slow and resource-intensive, making it essential to adopt data-driven methods. In this context, machine learning (ML) has emerged as a key factor for HEA research, offering the ability to efficiently explore compositional space and develop non-linear structure–property correlations. Classical approaches, such as tree-based ensemble models, have shown promising results and a domain-wide adoption. For example, Chang et al. applied ML algorithms (Random Forest Regression, Support Vector Regression, and Gaussian Process Regression) to BCC-based HEAs using features such as $\Delta H_{mix}$, the ductility (D) parameter, and $T_{test}/T_{melt}$, and found that the optimal model attained $R^2 \approx 0.90$ for predicting its yield strength [4]. Further analysis conducted in this paper explains that $\Delta H_{mix}$ and stacking fault energetics are among the most significant factors that control alloy strength. Qi et al. achieved over 90 % accuracy with Random Forest (RF) and Support Vector Machine (SVM) models to distinguish solid-solution vs. intermetallic phases in refractory HEAs [5]. ML-based approaches have also been used to predict properties beyond mechanical performance, like Zeng et al. employed an RF model to predict single-phase formability in AlCrCoNiFe systems [6]. With the goal of reliably identifying corrosion-resistant compositional regions, matching experiments. Although ML-based approaches have become increasingly popular, past work on predicting Vickers' hardness for alloy compositions remains limited. Recent work by Wang et al used a Random Forest framework to jointly predict phase structure, yield strength, and hardness [7]. This work achieved an $R^2 = 0.90$, RMSE = 48.91 HV while predicting Vickers' Hardness (HV). However, the prediction remained largely dependent on phase classification, and results did not generalize well across different phase regimes. While classical ML models was used to predict properties, two major problems were noted: (1) They rely heavily on handcrafted features (e.g thermodynamic descriptors, atomic properties, electronegativity difference between composition elements etc.) which increases dependency on data preprocessing and analysis (2) limited generalizability across broader compositional spaces as they often struggle to capture higher order, nonlinear interactions present in multi component alloys.

In these kinds of problematic situations, deep learning (DL) models offer a significant advantage over classical ML approaches by enabling automatic feature extraction in high-dimensional spaces, thus enhancing the modelling of complex nonlinear relationships common in multicomponent alloy systems. DL models like Graph Neural Networks (GNNs), in particular, have demonstrated strong performance in representing atomic interactions and molecular structures in chemistry and materials science [8]. Further, DL models like convolutional neural networks (CNNs), Bayesian-optimized deep neural networks, and transformer-based architectures leveraging multi-head attention mechanisms for both classification and regression tasks were also employed for predicting the properties of HEAs [9]. While these methods show promising accuracy and adaptability, a major challenge remains: DL architectures, which often consist of thousands to millions of trainable parameters, require large, diverse, and high-quality datasets to avoid overfitting and ensure generalization. Zhu et al. introduced a deep neural network trained on a dataset that one-hot encodes the manufacturing process (as-cast or laser AM) alongside elemental properties like atomic radius, electronegativity and thermodynamic metrics like mixing enthalpy ($\Delta H_{mix}$) to name a few [10]. The dataset used was supplemented with synthetic samples generated using a conditional generative adversarial network, and this additional training data resulted in a 27% drop in MAE. This highlights the importance of augmenting microstructural and thermodynamic property datasets for training DL architectures. Chaudhari et al. introduced AlloyBERT, a transformer-based language model that leverages Masked Language Modeling to learn composition–property relationships directly from natural-language descriptions of alloys

[11]. By pretraining on textual data of HEAs and fine-tuning a regression head, they achieved a mean squared error (MSE) of 0.00015 on the MPEA dataset and 0.00527 on the Refractory Alloy Yield Strength (RAYS) dataset. Although AlloyBERT demonstrates the promise of text-driven property prediction, its reliance on extensive pre-training limits applicability when large, domain-specific datasets are unavailable. Spyros Kamnis et al. also developed a transformer-based language model, using a generated pre-training dataset to predict elongation and ultimate tensile strength (UTS) for HEAs [12]. Although the generated dataset for language model pre-training and the curated dataset for property prediction were both highly diverse (aimed to improve generalizability), the resulting R² scores on the test set indicate space for improvement, achieving 0.561 for elongation prediction and 0.785 for UTS prediction.

Though, as stated, there are works that focus on predicting the hardness of HEAs, there are challenges with the limited data available in the domain. Hence, this work aims to address the challenge of predicting HEA properties under conditions of limited data availability and constrained computational resources. We employ pre-trained language models to infer Vickers hardness directly from textual alloy descriptors. By freezing the MatSci-BERT backbone and training only a regression layer, we achieve high accuracy with minimal data and computation. The framework further supports rapid virtual screening of HEAs, facilitating accelerated discovery of high-hardness alloys.

## 2. Methodology

The methodology of the present work consists of three main steps: data processing, model training, and predicting the hardness of the virtual data generated through combinatorics. The work employs the high-entropy alloy dataset reported in literature, which comprises 1,545 compositions of HEAs with corresponding hardness values used to train the ML models [13]. Since the aim of this work is to predict alloy hardness, those compositions that have null values present in the Vickers Hardness (HV) column are dropped, resulting in 415 unique HEA compositions.

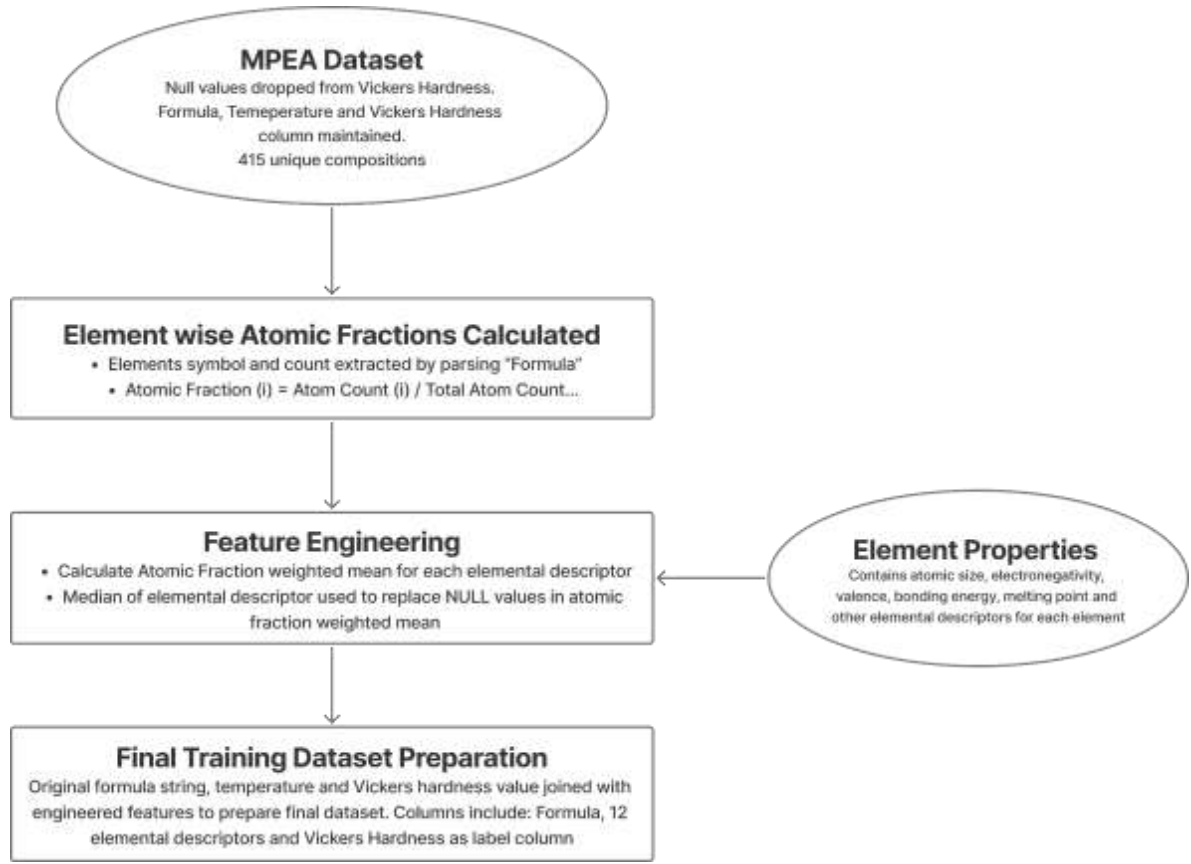

**Fig. 1.** Data cleaning and feature engineering flow chart.

*Fig. 1* illustrates the process of data preparation. Firstly, each composition is parsed and stored as a list of elements together with their corresponding atomic ratios. From this list, absolute atom counts are determined and normalized to atomic fractions. A curated elemental property table is then used to retrieve each element's 14 different features namely, Atomic Radius, Pauling Electronegativity, Number of Valence Electrons, Cohesive Energy (eV/atom), Bulk Modulus (GPa), Elastic Modulus (GPa), Shear Modulus (GPa), Melting Point (K), Rate of Shear Modulus

Change (MPa/K), Solid Solubility (at%), Lattice Constant (Å), Bond Electron Concentration (BEC/cm³), Average Valence Bond Strength (eV), and Engel's e/a ratio. Once the elemental properties are retrieved, these properties and corresponding atomic fractions were employed to feature the complete composition. These features are scaled using a standard scaler and partitioned into an 80–20 train and test split for training the different ML models.

The work employs three categories of models to investigate the prediction of HEA's hardness, namely, classical tree-based algorithms, transformer architectures with attention mechanisms, and language models. Transformer architectures and language models are employed to learn richer feature representations from scalar inputs, which are subsequently mapped to hardness values via a neural network. Alongside these, classical models are tested, as their simplicity and inductive bias align well with the dataset and feature design. The evaluation considers three classical models: Lasso Regression, Gradient Boosted Regression, and LightGBM, which enable mapping of scalar features to HV. To identify which elemental descriptors most strongly influence HV, a sparse linear model can be effective. Lasso Regression achieves this by reducing irrelevant feature weights to zero. The Lasso regression model minimizes the penalized least-squares objective, where the $L_1$ term enforces coefficient sparsity so that only the most relevant aggregated elemental descriptors (e.g., mean atomic radius, cohesive-energy variance) retain nonzero weights. The model is fit using a 5-fold cross-validation to select the optimal penalty strength, iterating up to 1,000 times with a learning rate of 0.1. This ensures stable convergence on the dataset and directly highlights which elemental properties most strongly govern HV. The gradient boosting algorithm is powerful in capturing nonlinear data. It builds the predictor as a sequence of decision trees, each one trying to reduce the deviation as compared to the previous ensemble. It starts from a simple initial guess (average HV), and each tree is then trained on the residual errors. By adding new trees in an ensemble manner, the model is gradually refined. A hyperparameter grid searches the number of estimators, tree depth, max features, minimum sample size, and minimum leaf size in a 5-fold cross-validation setup to assess generalization and select the optimal number of trees. The best-performing model parameters use the square root of the total number of features at each split and require a minimum of four samples per leaf and five samples to split a node. A fixed random seed during training helps ensure reproducibility. The LightGBM builds upon gradient boosting by the use of histogram-based binning and a leaf-wise tree growth strategy, which accelerates training and often yields more accurate splits on small datasets. A 5-fold cross-validation grid search algorithm searches over six hyperparameters i.e, number of estimators, max depth, number of leaves, learning rate, minimum child samples, and sub-sample ratio to identify the best possible configuration. The optimal model uses 100 trees of depth three with 31 leaves each, a learning rate of 0.1, subsampling ratios of 0.8 for both rows and features, and a minimum of five samples per leaf (trained with a fixed random seed for reproducibility).

Although classical models exhibit strong predictive performance, their limited capacity to extract features from high-dimensional spaces remains limited. DL architectures, like transformers by contrast, can automatically learn complex, hierarchical representations from raw data, reducing reliance on manual feature engineering due to their large number of trainable parameters that capture underlying complexity. Transformer models also use attention mechanisms, which allow them to capture global interactions amongst all input dimensions, effectively learning the data distribution and feature space. In the multi-headed attention mechanism for the HEA dataset, three types of pooling are employed: mean pooling, attention pooling, and CLS token. The overall architecture and training for all 3 models remain the same throughout our experiments to better evaluate results. The difference between the models lies in the final pooling of the 12 input vectors into a single summary vector for each alloy composition. This summary vector is then passed through a regression head with layer normalization and dropout to finally predict the HV of the HEAs. To train these transformer models, a batch size of 1 is employed; although computationally more expensive, this allows for faster convergence on small datasets. The model is set to train for 40 epochs with early stopping, which halts the training after 5 consecutive epochs when there has been no improvement in reducing the validation loss. Huber loss is used since it is robust to occasional outliers in hardness measurements and also behaves like mean squared error near the minima. AdamW optimizer is used with a weight decay of $1 \times 10^{-4}$ to prevent overfitting. ReduceLROnPlateau scheduler allows the reduction of learning rate when gradient descent is stuck in a local minimum, hence potentially improving the model capabilities by allowing model weights to converge closer to the global minima. The overall transformer architecture involves a linear embedding layer, which converts the 12 scalar input features of the alloy composition into a 16-dimensional vector. These enriched features are then passed through 2 stacked transformer encoder layers with 2 attention heads and a 128-dimensional feed-forward network, thereby capturing subtle dependencies and interactions between these features, which may affect the hardness of a composition. One of the 3 different types of pooling methods are now applied. (1) Mean pooling averages all token embeddings to create a single summary vector (2) Attention pooling uses the learned weights to compute a weighted sum, based on each feature's relevance (3) CLS token is a dedicated learnable embedding which is directly optimized during training and captures the global hardness prediction. This summary vector is then passed through the fully connected network working as a regression head to predict alloy hardness.

Language models use an encoder decoder architecture both built using one or many transformer blocks. Depending on the specific architecture these models are used for generation tasks (encoder-decoder models) and supervised learning (encoder only). To effectively utilise such powerful models with millions of parameters, we require a large training dataset. Due to our small training dataset (415 HEA compositions), a pre-trained MatSci BERT encoder is leveraged which is originally trained on millions of material descriptions. A transformer encoder–based model, inspired by Alloy BERT, is employed to predict composition hardness after training on custom dataset. The 3 strategies used include: (1) frozen-encoder transfer learning, (2) additional masked-language-model finetuning, and (3) skip-connection augmentation. In all cases, each alloy is represented as a concatenated string of its 12 compositional descriptors as prepared in the pre-processing step. This string is then tokenized and passed through the encoder, followed by a lightweight regression head. In the first strategy, MatSci BERT weights are loaded and frozen to preserve its learned chemical grammar (originally learnt using the large material science pretraining dataset). The final hidden state of the mean pooling layer feeds a five-layer fully connected regression head with ReLU activations. Only the fully connected head parameters are optimized during training using an AdamW optimizer at an initial learning rate of $1\times10^{-2}$. A layer-wise weight decay (0.02 for the first three linear layers, 0.01 elsewhere, and zero decay on all biases) is also applied to help the model regularize better. Training is performed for 100 epochs using an 80–20 split, with the learning rate reduced on a plateau to a minimum of $1\times10^{-7}$.

In our second strategy, we finetune the weights for the original MatSci BERT model upon recognizing that our HEA compositions and input strings differ from MatSci BERT's original training population. A 150 k-sample as provided by the AlloyBERT paper is used to perform masked language modeling in order to learn vocabulary and grammar specific to the alloy composition space. Tokens are masked with a 15 % probability, and the weights of MatSci BERT are fine-tuned for 40 epochs using a batch size of 16 and a learning rate of $5\times10^{-5}$. Similar to our first strategy, we freeze the encoder and retrain the identical regression head, yielding a significant reduction in validation MSE and improvement in $R^2$. Finally, the previous two strategies are extended by incorporating direct skip connections from the raw 12-dimensional compositional vector into the regression head. Refer to ***Fig. 3***, where the mean-pooled embedding from the final transformer layer is concatenated with the original scalar descriptors before the first linear layer, enabling the model to leverage both learned contextual representations and untransformed feature values. This hybrid architecture further improves Vickers hardness prediction by preserving fine-grained elemental ratios alongside the contextualized composition summary.

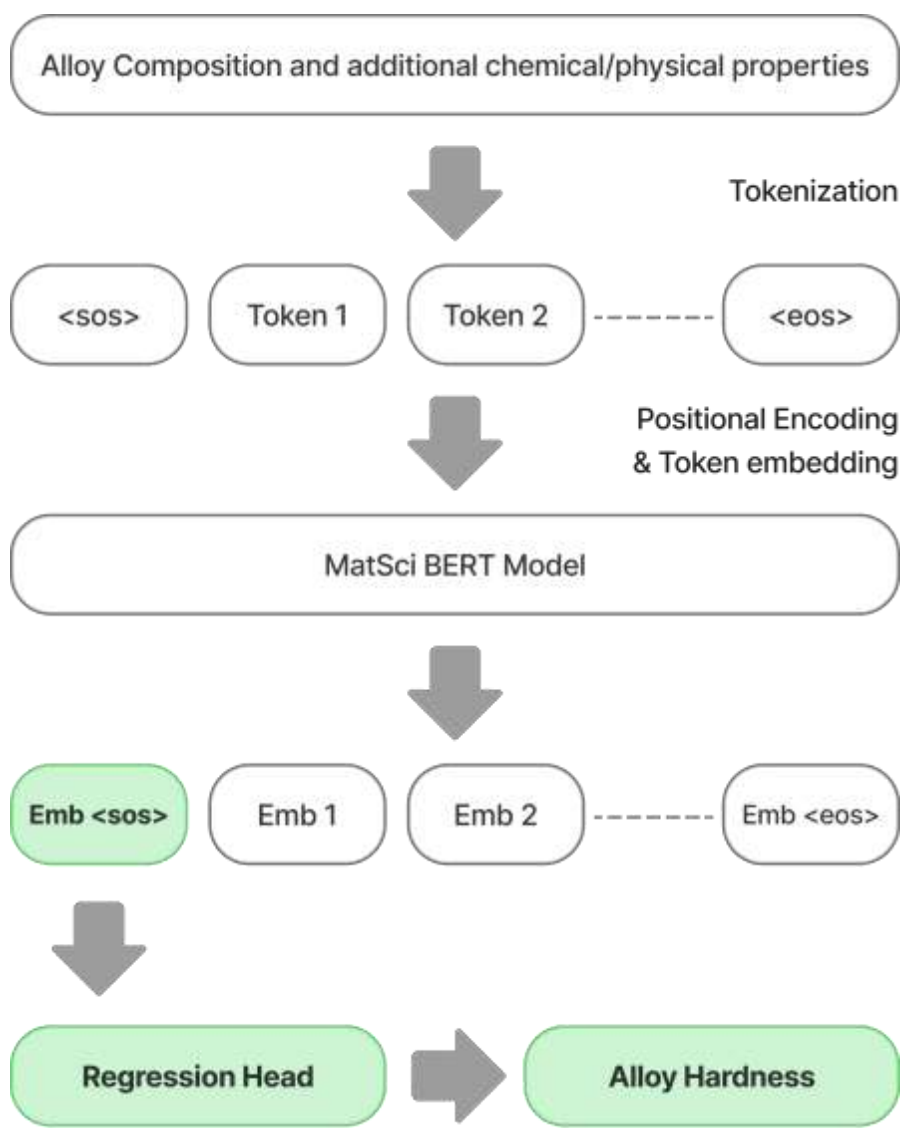

**Fig. 2.** BERT based encoder connected with regression head to predict hardness directly from natural language text.

In our second strategy we finetune the weights for the original MatSci BERT model upon recognizing that our HEA compositions and input strings differ from MatSci BERT's original training population. 150 k-sample as provided by the AlloyBERT paper is used to perform masked language modeling in order to learn vocabulary and grammar specific to the alloy composition space. Tokens are masked with a 15 % probability, and the weights of MatSci BERT are fine-tuned for 40 epochs using a batch size of 16 and a learning rate of $5 \times 10^{-5}$. Similar to our first strategy we freeze the encoder and retrain the identical regression head, yielding a significant reduction in validation MSE and improvement in $R^2$.

Finally, the previous two strategies are extended by incorporating direct skip connections from the raw 12-dimensional compositional vector into the regression head. Refer ***Fig. 3***, where the mean-pooled embedding from the final transformer layer is concatenated with the original scalar descriptors before the first linear layer, enabling the model to leverage both learned contextual representations and untransformed feature values. This hybrid architecture further improves Vickers hardness prediction by preserving fine-grained elemental ratios alongside the contextualized composition summary.

A virtual candidate library is algorithmically generated for all possible quaternary (primary element being Aluminium) alloy composition drawn from 14 secondary elements (Co, Fe, Ni, Si, Cr, Mn, Ti, Cu, Mo, Nb, V, Zr, Sn, Ta, Hf, W, Zn, Re, Mg, and Pd). Each secondary element is varied in 1 wt% % increments from 1 to 20. All possible weight-percentage combinations of the three selected elements are generated within their defined value ranges, with the remaining proportion assigned to Al to ensure a total composition of 100 %. The resulting alloys are expressed as structured composition strings (e.g., $Al_{68.4}Ni_{13.5}Zn_{13.4}Re_{4.7}$, $Al_{69.0}Fe_{10.9}Mo_{8.4}Zn_{11.8}$). This results in generating approximately 9.12 million virtual candidates. Rather than exhaustively evaluating each candidate, Bayesian Optimization is used to efficiently navigate the space. The trained language model (with skip connections) serves as the surrogate function, and uncertainty estimates are obtained by treating its outputs as probabilistic, using a Bayesian neural network approximation over the regression head weights. At each iteration, the surrogate returns both a predicted hardness and an associated uncertainty for every unevaluated composition. The next batch of candidates is selected by striking a balance between those with the highest predicted hardness (exploitation) and those about which the model is most uncertain (exploration). These chosen compositions are scored by the surrogate, the new data are incorporated to update its parameters, and the process repeats until we converge on a compact set of promising virtual alloys. This approach is inspired by the Bayesian optimization framework of Kristiadi et al [14].

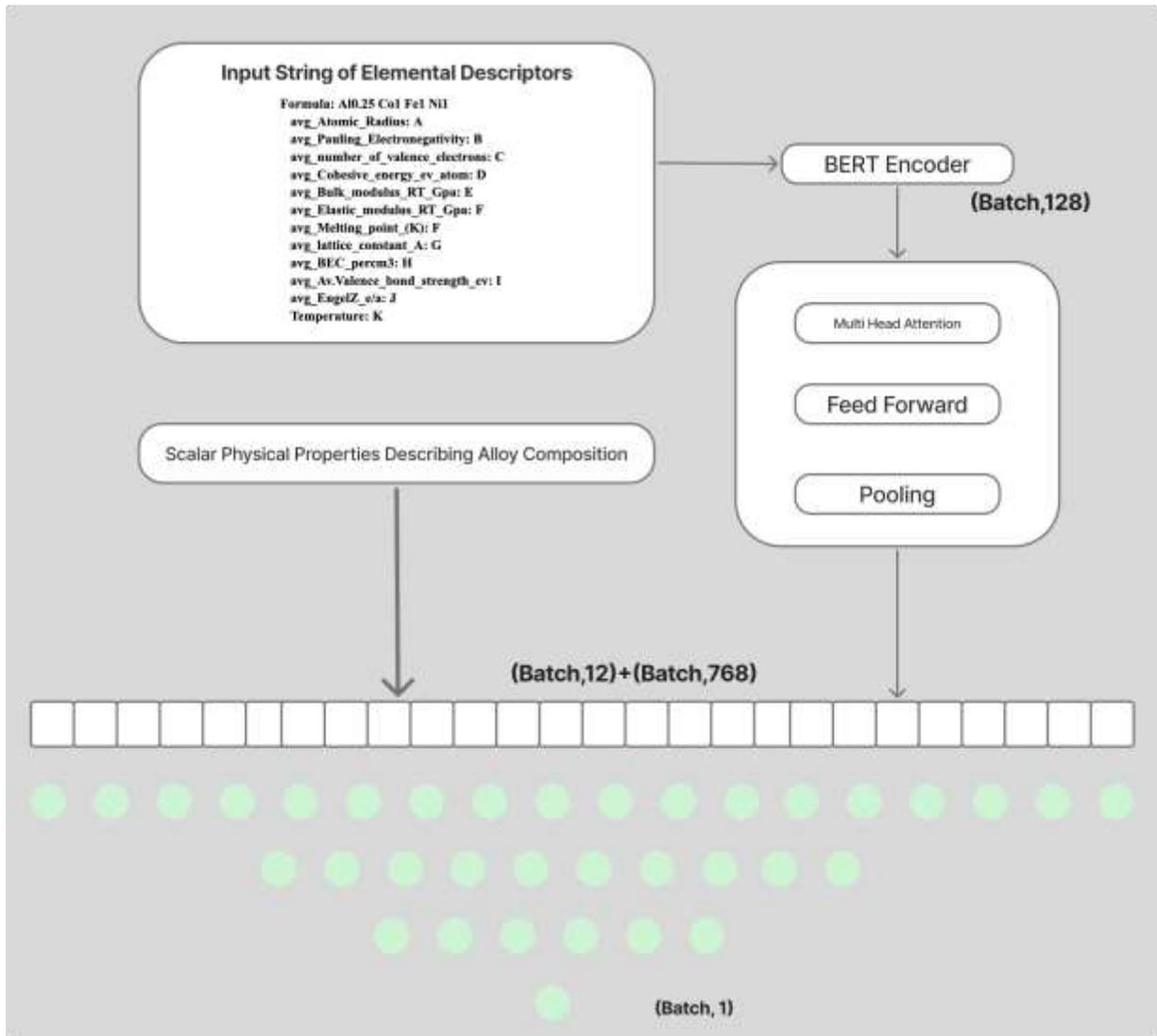

**Fig. 3.** Using neutral language alloy description along with scalar values for alloy descriptors (including temperature). Alloy hardness is predicted at the last layer.

## 3. Results and discussions

We study 3 different models to predict HV for HEAs. The results of the three classical regressors are present in *Table 1.* Amongst the classical models a steady improvement in predictive accuracy is observed with increasing model complexity. The Lasso regression model yields a test $R^2$ of 0.389 (MAE of 130.9 HV, RMSE of 26,956 HV). This confirms that sparse, linear combination of features is insufficient to capture the full compositional complexity. Introducing nonlinearity via Gradient Boosting improved generalizability on test set resulting in $R^2 = 0.748$, MAE = 78.0 HV, and RMSE = 11,141 HV. Further improvements were achieved with LightGBM (100 trees, maximum depth = 3, subsample ratio = 0.8), which reached $R^2 = 0.773$, MAE = 73.9 HV, and RMSE = 10,038 HV. LightGBM model effectively captured the data variability present in the MPEA dataset as it provided the best scores on the test set, and thus provided better generalization as compared to the other two models.

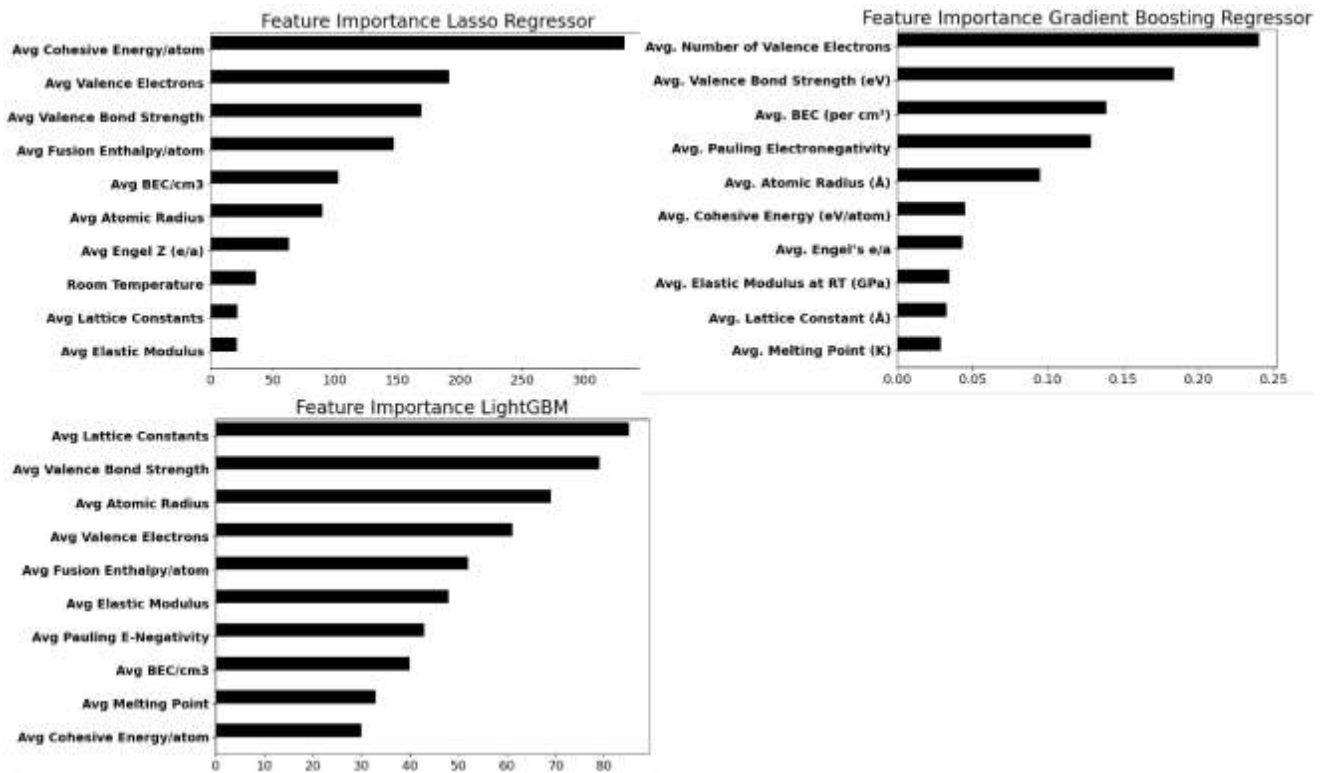

**Fig 4:** Top 10 elemental descriptors used to determine Vicker's hardness for the 3 classical models.

Table1: Training results for classical models.

| Model | Train R2 | Test R2 |
|---|---|---|
| Lasso Regression | 0.322 | 0.389 |
| Gradient Boost Regressor | 0.887 | 0.747 |
| Light GBM | 0.894 | 0.773 |

Feature importance analysis conducted for the top ten elemental descriptors identified by both the ensemble-based models and the Lasso regression model are presented in *Fig. 4*. In the Lasso model, cohesive energy appears most strongly, followed by average valence electron count, bond strength and fusion enthalpy. Both Gradient Boosting and LightGBM likewise rank valence electron count, bond strength, and lattice spacing in their top five, while atomic radius, cohesive energy, and shear-related properties also recur but at a lower importance. This consistency across the tree based models highlights the central role of electronic structure and bond characteristics in determining alloy hardness, and the coefficient of determination $R^2$ demonstrates superior performance and the advantage of ensemble methods in capturing their nonlinear interplay.

Table 2: Test set results comparing different Transformer pooling methods.

| Model | MAE | RMSE | R2 |
|---|---|---|---|
| Mean Pooling | 87.55 | 106.68 | 0.742 |
| Attention Pooling | 104.78 | 129.41 | 0.621 |
| CLS Token | 89.77 | 113.54 | 0.708 |

Building from the classical results, the three pooling strategies are evaluated in transformer-based regressors. The results on the test set are presented in *Table 2*. The model using simple mean pooling achieved the best balance of accuracy and generalization, with a mean absolute error of 87.6 HV, an RMSE of 106.7 HV, and an R² of 0.742 on the test set. *Fig. 5* illustrates training versus testing performance and loss progression for mean pooling transformers. The results show that a dedicated [CLS] token yielded slightly lower performance (MAE = 89.8 HV, RMSE = 113.5 HV, $R^2 = 0.708$), suggesting that a fixed–length summary embedding does not capture all of the compositional variance. The reduced performance of attention-weighted pooling (MAE = 104.8 HV, RMSE = 129.4 HV, $R^2 = 0.621$) on the test dataset is likely due to overfitting of the learned feature-based weights, evidenced by the highest training $R^2=0.84$ among all pooling strategies. These findings prove that, for small high-entropy alloy sets, straightforward aggregation can rival more complex attention schemes.

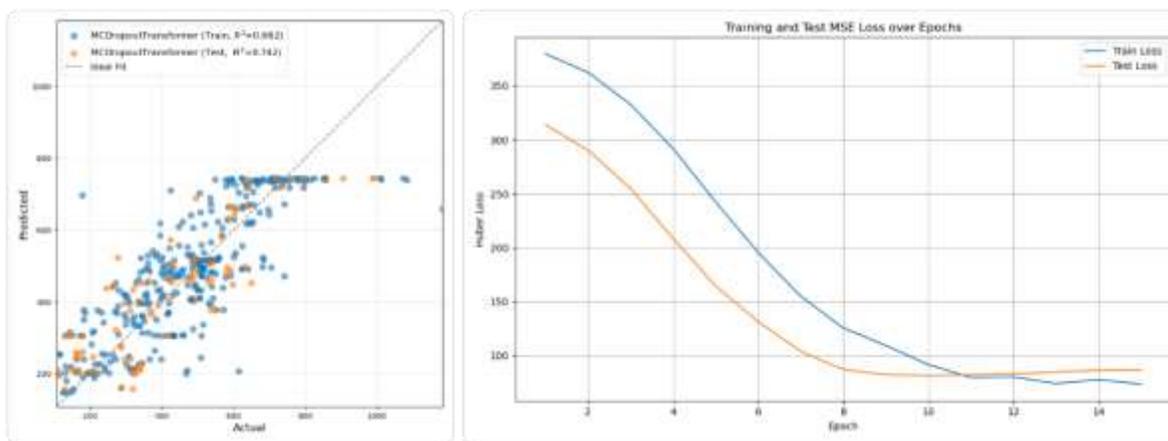

**Fig. 5.** (i) Predicted vs. actual Vickers hardness for the Mean Pooling Transformer on training ($R^2 = 0.662$, blue) and test ($R^2 = 0.742$, orange) sets; (ii) Training and test Huber loss curves over 15 epochs (early stopping).

The study extends beyond the training and evaluation of transformer-based regression models by employing a chemistry-aware language model (MatSciBERT) capable of capturing rich, domain-specific representations. Effective pre-training of BERT-style encoders requires a large dataset, which is often lacking in the material science domain, especially in the study of high entropy alloy compositions. Hence, MatSci BERT is utilized, which is trained on millions of material descriptions and learned attention patterns tailored to atomic interactions and molecular structure. This makes it an ideal backbone for encoding compositional strings in the HEA domain.

Table 3: Test results to compare 4 language model strategies.

| Model | MAE | RMSE | R2 |
|---|---|---|---|
| MatSci BERT (no skip connections) | 114.29 | 153.64 | 0.586 |
| Pre trained MatSci BERT (no skip connections) | 106.91 | 137.16 | 0.601 |
| MatSci BERT (with skip connections) | 94.1 | 118.82 | 0.672 |
| Pretrained MatSci BERT (with skip connections) | 83.4 | 102.67 | 0.762 |

Upon freezing the original MatSci BERT weights and training only a lightweight regression head, the model attains an MAE of 114.3 HV, RMSE of 153.64 HV, and $R^2$ of 0.586 (comparable to regression transformer blocks). Introducing masked-language-model fine-tuning on the 150k-sample HEA corpus sharpens its internal chemistry grammar, boosting performance to MAE = 106.9 HV, RMSE = 137.16 HV, and $R^2 = 0.601$. Finally, by concatenating the features extracted by the mean pooling layer with the raw twelve-descriptor vector via skip connections, the regression head recovers fine-grained elemental ratios alongside contextualized representations. This hybrid model achieves MAE = 83.4 HV, RMSE = 102.67 HV, and $R^2 = 0.762$, comparable to the top classical ensemble and surpassing all other transformer variants. Table 3 presents an overview of the test set metrics associated with each modeling approach described above.

To further validate the generalization capacity of our framework, we compared its predictions against an out-of-data composition with ThermoCalc measured hardness (**Table 4**). For the alloy $Al_{6.25}Cu_{18.75}Fe_{25}Co_{25}Ni_{25}$, previously reported by Singh et al., the model estimated 128.46 HV versus the ThermoCalc value of 150±10 HV [15]. This deviation highlights that while the model captures underlying composition–property trends, discrepancies arise from limited coverage of certain elemental interactions in the training corpus and potential microstructural factors (e.g., grain size, precipitate formation) absent in purely descriptor-driven representations. Addressing these gaps will require expanding the dataset with systematically measured HEAs, incorporating processing parameters, and exploring multi-scale representations that bridge atomistic descriptors with microstructure features.

**Table 4:** Predicted vs ThermoCalcHardness for Real HEA

| Formula | Predicted Hardness | ThermoCalcHardness |
|---|---|---|
| $Al_{6.25}Cu_{18.75}Fe_{25}Co_{25}Ni_{25}$ | 128.46 | 150 |

**Table 5:** Top 5 virtual candidates explored using Language Model and their EI scores

| Formula | EI | Std | Predicted HV |
|---|---|---|---|
| $Al_{57.8}Co_{13.2}Fe_{14.0}Cr_{15.0}$ | 3.30E-29 | 49.278 | 626.849 |
| $Al_{83.2}Mn_{5.3}Zn_{8.9}Re_{2.7}$ | 9.67E-39 | 47.825 | 558.263 |
| $Al_{85.3}Ni_{11.0}W_{0.6}Pd_{3.2}$ | 9.67E-39 | 47.913 | 557.104 |
| $Al_{80.5}Ni_{7.5}Mn_{8.6}Zr_{3.3}$ | 9.67E-39 | 48.207 | 553.242 |
| $Al_{85.7}Co_{3.7}Ni_{8.4}Si_{2.2}$ | 9.67E-39 | 47.997 | 556.003 |

Exploring the virtual candidate space (generated using combinatorics) using the language model and approximating predictive uncertainty using a Bayesian network on our regression head resulted in finding: $Al_{57.8}Co_{13.2}Fe_{14.0}Cr_{15.0}$ (626.849 HV), $Al_{61.4}Ni_{13.4}Mn_{13.6}Zn_{11.5}$ (570.8 HV), $Al_{61.4}Ni_{12.8}Mn_{14.4}Zn_{11.5}$ (571.5 HV), $Al_{61.5}Ni_{12.8}Mn_{13.7}Zn_{12.1}$ (555.5HV), and $Al_{61.4}Ni_{12.1}Mn_{14.4}Zn_{12.1}$ (556.2 HV). In particular, the Co–Fe–Cr system in the top candidate is known to form BCC or B2 phases that resist dislocation motion, while the Ni–Mn–Zn series benefits from significant atomic size mismatch and favorable mixing enthalpies to enhance hardness. Additionally, Table 5 depicts the moderate predictive standard deviations, which underscore the surrogate's calibrated uncertainty. This ensures that the BO loop judiciously balances exploitation of high-hardness regions with exploration of compositions where the model is less certain. Collectively, these outcomes highlight the ability of the guided approach to zero in on compositions exhibiting promising hardening mechanisms, thus offering a data-efficient route to down-select alloys for experimental screening using the language model.

4. Conclusion

In this study, a comprehensive framework showcasing data preparation, model training, and evaluation is adopted for predicting Vickers hardness of HEAs. Among classical regressors, LightGBM delivered the strongest performance ($R^2 = 0.773$, MAE = 73.9 HV), closely followed by Gradient Boosting ($R^2 = 0.748$, MAE = 78.0 HV). Simple transformer encoders with mean pooling achieved comparable accuracy ($R^2 = 0.742$, MAE = 87.6 HV), outperforming more complex pooling schemes due to the small size of the training dataset. Leveraging the MatSci BERT pre-trained weights further enhanced results. Masked-language-model fine-tuning and the addition of raw-descriptor skip connections, the frozen-encoder model matched the top classical ensemble ($R^2 = 0.762$, MAE = 83.4 HV) while maintaining a lightweight training footprint. The results of the language model regressor are verified by evaluating performance on an out-of-sample data set for which the HV value is calculated using the ThermoCalc software. A notable outcome from the self-attention visualizations was the consistent identification of atomic radius, valence electrons, and bond strength as the most influential features, indicating that the models learned importance measures align closely with established physical understanding. This pattern was observed not only in the language model but also in the feature importance rankings derived from classical models such as LightGBM and gradient boosted regression. Our results underscore two key insights: (1) ensemble methods remain highly competitive when the dataset size is limited. (2) Transfer learning with domain-pretrained language models can rival bespoke architectures at minimal additional cost. To further accelerate HEA discovery, future studies should integrate processing-parameter data (e.g., annealing temperature, cooling rate) into feature sets and explore graph-based neural networks to capture local microstructure properties. This would bridge the gap between physical descriptors and chemical interactions.